# Realistic estimate of the Covid-19 incidence and mortality rate in France

Guillaume MALPUECH[1*], Anne TOURNADRE[2,3], François BERRY[1], Laurent GERBAUD[1,4]

[1] Institut Pascal, CNRS, Université Clermont Auvergne, Sigma Clermont, F-63000 Clermont Ferrand, France

[2] Rheumatology department, CHU Clermont Ferrand, France

[3] Unité de Nutrition Humaine, INRAE, Université Clermont Auvergne, France

[4] Public Health Department, CHU Clermont Ferrand, France

**Abstract :**

Large scale virological testing of SARS-Cov2 is implemented since May 2020 in France. We assume that the positivity of asymptomatic people not being contact cases is representative of the positivity of the whole French population. This allows estimating the real incidence to be 0.8% on August $1^{st}$ and 2.4% at the beginning of September (1.6 million people, 230.000 daily infections). We deduce that intensive care units (ICU) admissions and infection fatality rates (IFR) dropped by one order of magnitude since March, and are currently 0.036% and 0.027% respectively. Basic simulations of the outbreak evolution assuming negligible reinfection probability are performed for France and different regions. These simulations are using an initial reproduction rate (R) ranging from 1.3 to 1.45 (1.15 in Grand Est) which then evolves only because of the growing immunity. An incidence peak of 3.5 % is expected at week 39 for France. The calculated total number of ICU admissions and deaths during the second wave are 9000 and 7000 respectively for R=1.3. The cumulative incidence over the two waves is computed close to 60% for R=1.3. This suggests that if individual immunity exists, herd immunity could be achieved in France during the autumn 2020.

*Corresponding author: guillaume.malpuech@uca.fr

**I Introduction**

To predict the Covid-19 outbreak trajectory and adjust public policies, the knowledge of the total (cumulative) infection rate and the instantaneous infection rate (also called incidence) of a given population is crucial. Since May, large-scale virological testing is executed in France. About 300.000 tests a week were performed in May, a number which is reaching 1 million at the end of August and beginning of September. The positivity of these tests was around 2% in May, fell to 1% in June and then continuously rose to reach 5.2% on September 9$^{th}$. At first glance, it seems plausible that these numbers tell something about the real fraction of the population infected. However, this testing was neither conceived nor currently analyzed in a way that could provide such information. The idea in May was that the outbreak was not spread, and that to control it one needed to find clusters. The hypothesis made by authorities was that most of the infected people are really found by the tests. The quantity which is put forward and used by the French authorities is the total number of positive tests wrongly called incidence. This number of positive tests is still used as a key parameter to determine the status of a department. In that vision, the fraction of positive tests, also called "positivity" had not much meaning, because the tests were targeted. They were centered at people likely to have been infected who form the clusters. However, at least in July-August, the tests performed are not linked with clusters or to "contact-cases" in their wide majority. A majority of tests is performed on asymptomatic people not being contact cases (ANBC). It is therefore highly plausible that the positivity measured on this population (ANBC) *is* a good measure of the real incidence, namely the fraction of the population infected at a given time. Indeed, ANBC can be considered as non-correlated data-based sampling of positive cases.

**II Determination of the real Covid-19 incidence in France.**

Our goal is therefore to compute the positivity of ANBC which we judge a reasonable measure of the positivity in the general population. Positivity of asymptomatic can be either straightforwardly computed, or is directly provided in the reports provided each week by Santé Public France. In order to extract the positivity of ANBC which is not directly provided in the reports, one has to remove the contribution from contact cases. To be more quantitative, we consider the report of Santé Public France of August 13$^{th}$ [1] presenting data for the week 32. Asymptomatic persons represent 77 % of the 540.000 persons tested in total (so 387.800 persons). Asymptomatics represented 53 % of positive tests (page 6). The positivity of asymptomatics is therefore 0.53/0.77=0.69 of the total measured positivity (P) which is (2.2%). This gives a positivity of asymptomatic's of 0.69xP=1.52% and 6320 positive tests. On the other hand 28000 contact cases were identified on week 31 and 2336 were found positive which corresponds to a positivity of 8.3 %. Then assuming that these contact cases were all asymptomatic allows to find the number ANBC tested (0.77x540.000-28.000 = 387.800). The number of positive test among ANBC is therefore 6320-2336 = 3984. The positivity of ANBC was therefore at least 3984/387800 = 1.03 %, which makes (0.47xP). This is a minimal value since there is some probability that a part of contact cases then tested positive happened to be symptomatic and did not count in the positive of asymptomatic. The ratio between symptomatics and asymptomatics is not very well known and many different estimates can be found in the literature. The results of this manuscript suggest that the fraction of asymptomatics is probably very large, and we therefore assume a value of 90 % which brings only a small correction to the positivity of ANBC which is reaching 1.09 % (0.5xP) using this procedure. Table 1 presents for weeks 32, 36 [2] and 38 [3], and 40 [4] the positivity of all tests, the positivity of symptomatic and asymptomatic people tested, the minimal and most plausible positivity of ANBC evaluated as above.

Table 1: For weeks 32,36,38,40 positivity of all tests, of Symptomatic, Asymptomatic, Minimal and most plausible positivity of ANBC computed as explained in the text. Evaluated number of people infected assuming that the incidence in the general population is similar to the one of tested ANBC.

| Week | Positivity (P) | P of Symptomatic | P of Asymptomatic | Minimal P of ANBC | Most plausible P of ANBC | Population infected (Incidence) |
|---|---|---|---|---|---|---|
| 32 | P=2.2% | 2.05 P=4.51% | 0.69 P=1.52% | 0.47 P=1.03% | 0.5 P= 1.1% | 737.000 |
| 36 | P=5.2% | 2.23 P=11.6% | 0.62 P=3.2% | 0.46 P=2.37% | 0.47 P=2.45% | 1.641.000 |
| 38 | P=6.2% | 1.53P=9.5% | 0.63 P=3.9% | 0.4 P=2.46 % | 0.46 P=2.6% | 1.742.000 |
| 40 | P=9.1% | 1.81 P=16.5% | 0.59P=5.4% | 0.31P=2.85% | 0.38P=3.45% | 2.310.000 |

One can notice that between week 32 and week 36, the minimal positivity of ANBC was multiplied by 2.3. Assuming a generation interval of one week, and a constant reproduction rate R over time allows to estimate a minimal R-value of 1.23 (1.23^4=2.29). This is a minimal value since it neglects the decrease of the susceptible population over time. We do not compute any uncertainties on these numbers. The statistical uncertainty is extremely small regarding the large samples tested. It is much smaller that all the rounding performed to present the number. Uncertainties are entirely contained in the hypothesis that ANBC positivity is representative of the positivity in the general population. How much this is wrong or correct is very difficult to estimate. Our view is that this is a quite reasonable assumption that we will then assume to be fully verified. So considering the 67 million inhabitants in France, 2.45 % of positivity at week 36 makes **about 1.64 million people infected.** Assuming that a PCR test stays positive in average 7 days, allows to deduce **about 235.000 new infections each day.** This is the first straightforward result of this manuscript. This of course could be more rigorously evaluated by testing trully random samples.

Another conclusion is that **the testing finds typically 5 % of infected people**, because simply it is performed over only 2% of the full population each week, a conclusion was partly by other groups in France[5]. There are many consequences of this simple fact. It means that about 1 person over 40 is infected (1 over 20 in Marseille or Paris). It means that there is almost one kid or one teenager infected in each class of France, and for sure several ones in each school. It means that closing a class just because one kid was tested positive has a very little meaning, which has indeed been acknowledged by the government since three positive tests but one in a class are now require to close it.

**III Determination of the real mortality rate and the ICU admissions in France during the second wave.**

The mortality rate and the ICU admission probability in case of infection show a very strong age dependency. In fact, only people above 65 years old significantly contribute to mortality and ICU admissions. Determining a generic Infection Fatality Rate (IFR) for a complete population do not necessarily have a very strong meaning. Despite that, since June at least, there is a reasonable correlation between the ICU admissions and the positivity of tests. The number of weekly ICU admissions reached a minimum around 70 in June and rose to almost 300 the first week of September [2], which makes a factor 4-5. During the same period, the positivity of tests rose from 1% to about 5%. We keep here a two weeks shift for comparing the ICU entrance and the test results. So, it seems that these quantities are correlated and increased by a factor 4 to 5 since June. Regarding death rate, a shift of three weeks between infection and death is commonly accepted. The table 2 summarizes the number used to compute the ICU admission rate and the IFR corresponding to the infections which occurred place on weeks 32 and 36 respectively.

Table 2: For weeks 32 and 36, incidence estimated in this work, ICU admission two weeks later and decuced ICU admission per infection. Number of deaths three weeks later and the deduced IFR. The two IFR values for weeks 36 are with and without including nursing home death.

| Week | Incidence estimated in this work | ICU admission two weeks later | ICU admission per infection | Death (Hospital) 3 weeks later | Death (including nursing home) 3 weeks later | Infection Fatality Rate |
|---|---|---|---|---|---|---|
| 32 | 576.000 | 174 | 0.03% | 109 | 109 | 0.019% |
| 36 | 1.641.000 | 599 | 0.036% | 336 | 452 | 0.02%-0.027% |

The ICU admission rate per infection is 0.03% for week 32 and 0.036% for week 36.

There was no declared death in nursing homes on week 32 (in fact the number declared is negative and we did not consider it). The IFR computed with hospital death only is 0.019% and 0.02% respectively. It is rising to 0.027 % considering death in nursing home on week 36.

Taking this rate and considering the 30.000 people who died during the first wave in France would mean more than 100 million people infected, which is of course not possible.
- One explanation is that the IFR here computed is wrong.
- The other option is **that the mortality enormously dropped between the first and second wave, by probably around one order of magnitude.**

Estimations of the IFR during the first wave of the outbreak in Europe gives values of 0.3-1.3 % [6,7]. As we will discuss in the fourth section dedicated to the first wave description, we evaluate, that a correct value for France is the lower bound of this interval (0.3%). In order to support the plausibility of the computed IFR, we first put forward outbreak data from a neighboring country of France which is Luxembourg (650.000 inhabitants). Luxembourg deployed a large scale testing policy since the end of June and is testing about 5-10 percent of its population each week [8]. For instance, 36.821 tests were performed on week 39 with a positivity of 1.25%. 14 persons died between July 11[th] and August 17[th]. The next death occurred on October 2[nd]. It means that there was no death recorded in Luxembourg during 44 days. During the same period 2200 people were tested positive. If we shift back by three weeks the period over which we count the number of positive test, we reach a value of 1800. It means that the probability not to die being infected is at least 99.94 %. With a probability 99.55%, the IFR in Luxembourg is below 0.3 %. With a probability 50 %, it is below 0.038%. We insist that this evaluation is about the full population of a country, including old people. Only a part of the population is tested, so the number of contaminated people is certainly larger and probably much larger than 1800. This evaluation is not a proof, but it is extremely compatible with the number found in the present manuscript. We stress that between the end of June and beginning of August the positivity and the number of positive tests were quite similar to what they were later in August and September. On the other hand, the number of dying people passed from 14 during 5 weeks, to 0-1 after. This is compatible with the fact that after a wave of the outbreak, IFR drops by a large amount. Another example comes from a recent study performed in Japan [9] where an astonishing rise of seroprevalence was measured during the summer among companies workers in Tokyo, reaching values close to 50 %. In the same time, the number of death in the country remained extremely low (500 people died of covid) during summer in the full Japan. The last example comes from a preprint where the authors compared the evolution of patient hospitalized in their hospital during the first and second wave [10] of the outbreak in France. They argue to use the same criteria to hospitalize people during the two waves. They found a decay by a factor 5 of the mortality of the patients admitted to hospital during the second wave with respect to the first wave. If one combines this factor with a decay in the probability to be hospitalized while infected, one can easily reach a factor comparable with the one we find in our work.

Several explanations can be put forward to explain this drop.

- The first one is evidently the improvement in disease management and care of critically ill patients.
- A second one much more speculative regards the often mentioned younger mean age of infected people. In the absence of any systematic record during the first wave in France, this hypothesis cannot be verified. The one order of magnitude decay we find would require a decrease by more than 20 years of the mean age of infected people. This explanation could play a role but is unlikely to fully explain such a change. Indeed, the incidence on elderly people is all along this second wave typically 2-3 times smaller than the one if the 20-29 years old population. First, this factor is not an order of magnitude. Second, young people do have more social contact than elderly people. It was true in March 2020 when Universities, bars and night clubs were opened. It is still true now. Elderly people have reduced their interactions, but young people as well. This change in the age distribution of infected people is therefore not proven and is not, in our opinion able to explain a very large IFR.
- Another possibility is a weaker activity of the virus on infected people, which can be due to unknown factors. One can cite a possible mutation of the virus or a development of immunity in the population, not necessarily against infection, but against the development of symptoms. The wearing of masks and social distancing may limit the quantity of virus received by a person while he gets infected. Influence of the season might also be evoked. It remains that the drop of mortality is a mathematical fact, whatever its exact origin.

**IV Cumulative incidence during the first wave.**

In this section we estimate the cumulative incidence during the first wave of the outbreak in France. Very little is known about the early evaluation of the outbreak in January-Ferbruary [11]. Then, very few virological tests were available in March-April and the real incidence during this period is not very well known. Two influential sources in France estimating this value are a numerical simulation of outbreak evolution published in Science [6], which found a total infection rate of 5.3 %, and, more recently, a serological measurement published by Sante Public France on July 9th, reporting a rate of 6.73 % [12] IC (5.36-8.11). This last testing was performed during the week 6th-12th April. One should notice that the antibody rate decays versus time after the infection, as now well documented [9, ,13,14,15,16]. Not surprisingly, this also happened in France. The Santé Public France seroprevalence study [17] for the week 20 (May 11th-17th) found 4.93 % (IC 4.02-5.89) of positive test, a value being out of the confidence interval of the measurement performed 5 weeks before. This decay explains why everywhere these tests are showing relatively low values. More precisely, they rise at the heart of the outbreak then peak and probably decay. Serological tests results do not follow the dynamics of other signals such as hospitalizations and ICU admissions, may disappear after 90 days or even be undetectable in asymptomatic or mild forms of the COVID-19. Therefore, they should rather be understood as a delayed instantaneous signal rather than a measurement of cumulative incidence. If we take the case of France, the measurement performed on April 6th-12th is a minimum value of the fraction of people infected before March 26th (to let time to the antibody response to grow). Then, we analyze the number of hospitalizations and ICU admissions related to the first wave, so occurring before June 1st. We separate these quantities in two. Those which occurred before April 2nd included, and those occurring after (so in April-May). The number of ICU admissions is the same for the two periods. The number of hospitalizations which occurred after April 2nd is 1.37 times larger than the one which occurred in March [18]. It is therefore logical to assume that the real cumulative incidence is between 2 and 2.37 times 6.73 %, which is already a lower bound. This yields cumulative incidence between 13.4 and 15.95 %, which can be approximated to 15% for simplicity, which makes 10 million people. This is moreover assuming that the severity of infection remained constant, whereas as discussed previously, this severity appears to have strongly decayed with time. Then, comparing mortality rates, one gets that the Ile de France region gets an infection rate 2.25 times larger than the mean value in France, which makes a cumulative infection rate of 33 %. Finally, for the period June-July, we estimate that the ratio between hospital/ICU admissions and number of infection already dropped with respect to March. We therefore estimate the cumulative incidence using the positivity of tests which became large scale at that period. This allows to estimate a cumulative incidence around 5% for this period for France. Indeed, in June-July, the positivity of test was always between 1 and 1.5%. We found in the previous section that the positivity of ANBC is about 0.45 of the total positivity, which gives positivity of ANBC around 0.45 to 0.65% which multiplied by the 9 weeks of the period June-July indeed gives around 5%. Such estimates give a quite larger value of 10 % for Ile de France. One could also notice that the mortality during the first wave can be evaluated to 0.33 % (30.000 deaths, 10 millions infections) which is compatible with the lower bound estimates of IFR [7], but 12 times more than the IFR we estimate in France for the second wave. Therefore, the estimated cumulative incidence for the period February-May which can be called the "first wave" is 15%. The cumulative incidence over the period June-July which could be called the beginning of the second wave is estimated to be around 5%. **Therefore, the cumulative incidence at the end of July is estimated, for France around 20 % (and 40% for Ile de France).**

**V What about the future evolution of the outbreak simulated with a constant $R$.**

In this section, we present a few basic simulations of the outbreak evolution using the numbers found in the previous section and a few hypotheses. One key hypothesis is that re-infection cannot occur in a significant way. This is fully speculative, but could be checked experimentally easily, as commented in the discussion section. The second hypothesis is the estimate of the viral transmissibility $R$ which is the average number of people who will be infected by a person prior to isolation. Of course, this factor depends on many parameters, such as social distancing, restrictions, fraction of susceptible people, etc...

We first assume that $R$ is constant over time. Indeed, restriction policy remained relatively constant since the end of July, and only starts at the very end of September to evolve with the complete closure of bars in some limited area, so far. On the other end of summer vacations at the end of august, sudden change of weather which became much colder at the end of September do change the propagation condition for the virus and are impacting. We let such more precise evaluations for another forthcoming work. We therefore use a time-independent $R$-value that we estimate as good as we can from the growth of positivity, of hospital/ICU admissions. The rise of positivity of ANBC computed over 4 weeks in a previous section corresponds to a minimum R-Value of 1.3. The maximum rise of positivity from one week to another reached 40 % at the beginning of August. Regarding the ICU admissions, the maximum increase from one week to another was of 45 % between week 37 and 38. So, our simulations are using an effective reproduction rates ($R$) for France of 1.3, 1.4 and 1.45. We would like to stress that $R$ of the order of 3 was found in March, which allowed to deduce a "herd immunity" threshold of $1-1/R$ of 66 % of a susceptible population widely quoted in mass media. Whatever the reason which reduced the $R$ value, $R$=1.3 means a herd immunity threshold of 23 % of the susceptible population and $R$=1.45 means a herd immunity threshold of 31 %. Positivity started to rise at the end of July (week 30), which is the *t=0* of our simulation for France. We take 0.6 % of real incidence at that time. Despite several recent publications [19,20], we assume no cross immunity because it would lead to too many uncontrolled assumptions. On the other hand, we consider that a non-zero fraction of the population was previously infected and could not be infected a second time. The computation of this quantity is explained in the previous section. We take 20 % for France at week 30, 40% for Paris. We qualitatively estimate that these values could be 20 % for Bouches-du-Rhone, 10% for Puy de Dome which was weakly affected during the first wave. An homogeneous propagation model is used for the evaluation of the virus spreading. This is clearly very over-simplified. In reality the propagation is very inhomogeneous both spatially and from individual to individual. The fact that contaminations occurs mainly through super-spreaders and is very stochastic is a key element of this outbreak [21,22].On the other stochasticity driven by super spreading events is expected to play a strong role when the outbreak is emerging. The situation during the second wave is much more homogeneous, with a wide spreading which makes the description in terms of homogeneous mean values an acceptable approximation we believe. The proposed model is recursive and based on the following data:

- $I_0$ is the initial incidence (0.6 % for France at week 30)
- $f_s$ the susceptible fraction of the population when the simulation starts

$$N_{i+1} = N_i R \left(1 - \frac{I_0}{f_s} \sum_{k=0}^{i} N_k \right)$$

At the start of the simulation (t=0) $N_0 = 1$ and $N_0 I_0$ is the initial incidence at this moment. The term $\sum_{k=0}^{i} N_k$ is at the week $i$ the total number of people infected by 1 person infected at t=0. The susceptible population at the week $i$ is $f_s - I_0 \sum_{k=0}^{i} N_k$.

We perform simulations for France (full country) and for different regions. All parameters used are summarized in the table 3. Figure 1-a shows the computed incidence for France for two different $R$. The curve computed with $R$ = 1.3 gives an incidence close to 3% at week 37, which is in agreement with the experimental value. **This curve for France shows a peak arising at week 40 (end of September) with a maximal incidence of 3.5 %.** To establish a correspondence with the number of ICU admission per week, we use the larger value obtained in the section 3 which is 0.036%. With these parameters, the number of ICU admissions per week should go above 700. It could reach 1300 if $R$ is closer to 1.4. Figure 1.b shows the cumulative incidence and ICU admissions during the second wave. With $R$ =1.3, about 40 % of the population would be infected, and there should be around 9000 ICU admissions spread over several months and 7200 deaths using the larger IFR we computed (0.027%). The cumulative incidence including both the first and second wave would be 60 %. This is large enough to avoid a large third wave even with fully relaxed social distancing. Of course this herd immunity can be achieved only if individual immunity does exist. These simulations constantly use worst case parameters and the reality might be a bit more favorable by many aspects. Fraction of the population infected at the end of July might be a bit larger than the 20 % estimated. Cross immunity may indeed be non-negligible. IFR, and severity might suddenly drop further as it happened in Luxembourg.

Figure 2.a shows a comparison between the calculated incidence shown on fig. 1a and the hospital admission rate in France which stopped to increase at the end of week 39 and beginning of week 40. The hospital admission is shifted by one week (in fact a bit less), and the curve showing incidences are rescaled, just to be able to compare the shape of the curves. The agreement is satisfactory. The crucial point will be to know if the slight decay of hospital admission will sign the peak of the second wave, or will just be some statistical fluctuation. The shape of the experimental curve could be better reproduced by freely playing with parameters. The agreement between the simulation and the hospital admission rate is visually improved increasing both $R$ and $f_s$ and by shifting the hospital admission curve by 2 weeks instead of 1. The result is shown on figure 2.b. With a lot of caution, it could indicate that we overestimate the susceptible population, may be because of the neglected cross-immunity.

We then realized simulations which aim to describe the situation in three different regions of France, which are Ile de France, Bouches du Rhones, and Puy de Dome and Grand Est. Table 3 shows the parameters used.

**Table 3.** Table shows, for France and different regions the values of $R$ at the end of July, the values of $R_f$ which is $R$ if $f_s$ would be 100%, the estimated value of $R$ in March ($R_0$), the time t=0 when simulations start, the incidence $I_0$ and the susceptible population $f_s$ at t=0. When two value are shown the one in bold is the most likely.

| | Population (Millions) | $R$ (end of July) | | $R_f$. $R$ if $f_s$ would be 100% | $R_0$ (March) | t=0 | $I_0$ | $f_s$ |
|---|---|---|---|---|---|---|---|---|
| France (Fig 1,2a) | 67 | **1.3**-1.4 | | **1.625**-1.75 | 3-3.8 | Week 30 | 0.6 % | 80 % |
| France Fit of hospit. (Fig 2b) | 67 | 1.4 | | 1.75 | | Week 30 | 0.6 % | 60 % |
| Ile de France (Fig .3) | 12 | 1.3-**1.4** | | 2.17-**2.33** | 3-3.8 | Week 30 | 0.8 % | 60 % |
| Puy de Dome (Fig. 4) | 0.65 | 1.3 | | 1.44 | | Week 32 | 0.6% | 90 % |
| Bouches du Rhone (Fig. 5) | 2 | 1.45 | | 1.81 | | Week 30 | 0.6 % | 80 % |
| Grand Est | 5.5 | $R_f \times f_s$ 0.86-1.09 | 1.15 | 1.44-1.81 | 2.8 | Week 32 | 0.6% | 60 % |

Figure 3 shows the simulations for Ile de France. A difference with France is the estimated incidence at the end of July which is larger. At week 30, the positivity of test was 1.8 % which gives an incidence of ANBC close to 0.8 %. The positivity in Ile de France over week 36 is around 7 % which corresponds to an incidence of ANBC of 3 %. This is better reproduced by taking $R$ =1.4 instead of 1.3. It is not extremely surprising that the reproduction rate in Paris, even strongly modulated by pre-existing immunity, could be larger than an average for France. If only 20 % of the population would be immune in Ile de France instead of 40 %, the $R$ should be as large as 1.86. With no pre-existing immunity $R$ would be 2.33. It is difficult to estimate the precise value of $R$ in March. ICU admissions increased in Paris by a factor 3.8 between March 18[th] and 25[th] of and of a factor 3 between March 25[th] and April 2[th]. The difference between this 3-3.8 values and the "$R$ if $f_s$ would be 100%" in July (2.3) can be associated to the social distancing and restrictions. The decrease of $R$ from 2.3 to 1.4 in on the other hand related to the development of immunity. If we keep the value 3.8, the $R$-reduction due to the social distancing/restrictions and immunity respectively was in Ile de France of the same order at the end of July. With the growth of the 2[nd] wave, the role of immunity is gradually becoming dominant. **The peak is expected to occur for Ile de France at week 38.** The cumulative incidence for the second wave is about 40%, which combined with the estimated 40 % at week 30, gives a total fraction of population infected of 80 % in Ile de France at the end of October. Figure 3 shows a simulation performed for Puy de Dome. Puy de Dome is probably representative of an average for France in terms of urban density, but it is geographically quite isolated and has been weakly affected by the first wave. We therefore consider a preexisting cumulative incidence of only 10% and $R$=1.3. It means that the $R$ with 100 % susceptible population would be only 1.44. This is much smaller than in Ile de France which is reasonable considering the very different population density in the two regions. The second wave of the outbreak really emerged two weeks later than in the rest of France and we take a t=0 on week 32. As a result, the peak occurs only on week 43 (end of October) and the incidence could remain significant till the end of the year. The ICU admission could reach 7 per week, 70 admissions could occur in total. On the other hand one again should be very cautious since we are touching the limit of the basic model we are using which neglects any coupling between different areas [23]. It not clear that the outbreak in Puy de Dome has a strong dynamics by itself. It was possibly strongly driven by exchanges occurring during at the end of the holydays. The "local" reproduction rate might therefore be smaller than the one we considered. Figure 4 shows the simulation for the last considered area which is Bouches du Rhone, which cumulate a "large" reproduction rate being a large urban area (we take $R$=1.45), but was quite preserved during the first wave. The simulation suggests that the peak could be achieved at week 39, with an incidence of almost 7 % (more than 10 % for test positivity) and 45 ICU admissions a week. The last data suggests that the peak happened in Bouches du Rhone around September 15$^{th}$.

We finally discuss the case of the Grand Est region. Grand Est was the first region affected in France [11]. The mortality per inhabitant during the first wave was comparable to the one of Ile de France, and one can therefore estimate a comparable cumulative incidence for this period (40%). On the other hand the second wave appears to be quite limited. We can tentatively explain the difference between regions by the population density in Grand Est which is significantly smaller than in Ile de France. One can therefore expect therefore a smaller reproduction rate. The fact that the cumulative incidence of the first is comparable in both regions despite the different $R$ is due to the earlier start in Grand Est with respect to the lockdown which occurred on March 17$^{th}$ for all regions. The $R$-value in July, if $f_s$ would be 100 % in Grand Est is therefore expected to be significantly smaller than in Ile de France, but comparable with the values we evaluate for the rest of France which are in the range 1.44-1.81. Then considering the $f_s$ of Grand Est at 60%, we could deduce a $R$-value at the end of July in Grand Est to be in the range 0.86-1.08 as shown in the table 3. If we look at the Sante Public France Report for this region [24], It does not allow to compute the positivity of ANBC, but only the total positivity of tests which rose by a factor 6 between week 28 and week 39. Assuming a pure exponential growth yields $R$=1.18. Hospital admissions grew by a factor 3 between week 32 and week 39 which would give $R$=1.13. This value is slightly out of the interval 0.86-1.08 but remains close which, regarding all approximations in these evaluations remains quite satisfactory. The positivity of tests in Grand Est reached 1.4 at week 32. Using a ratio 0.4 between the positivity of ANBC and the total positivity found in other sections, we take $I_0$=0.6% at week 32 for the start of the simulation. The result is shown on Fig .6. A very moderate peak is found on weeks 40-41, with less than 30 ICU admissions a week which for the Grand Est region is relatively modest. The cumulative incidence over the second wave is slightly smaller than 20 %, which is less than for the first wave. After the two waves, the cumulative incidence will be of the order of 60 %, which is similar to the other French regions except Ile de France.

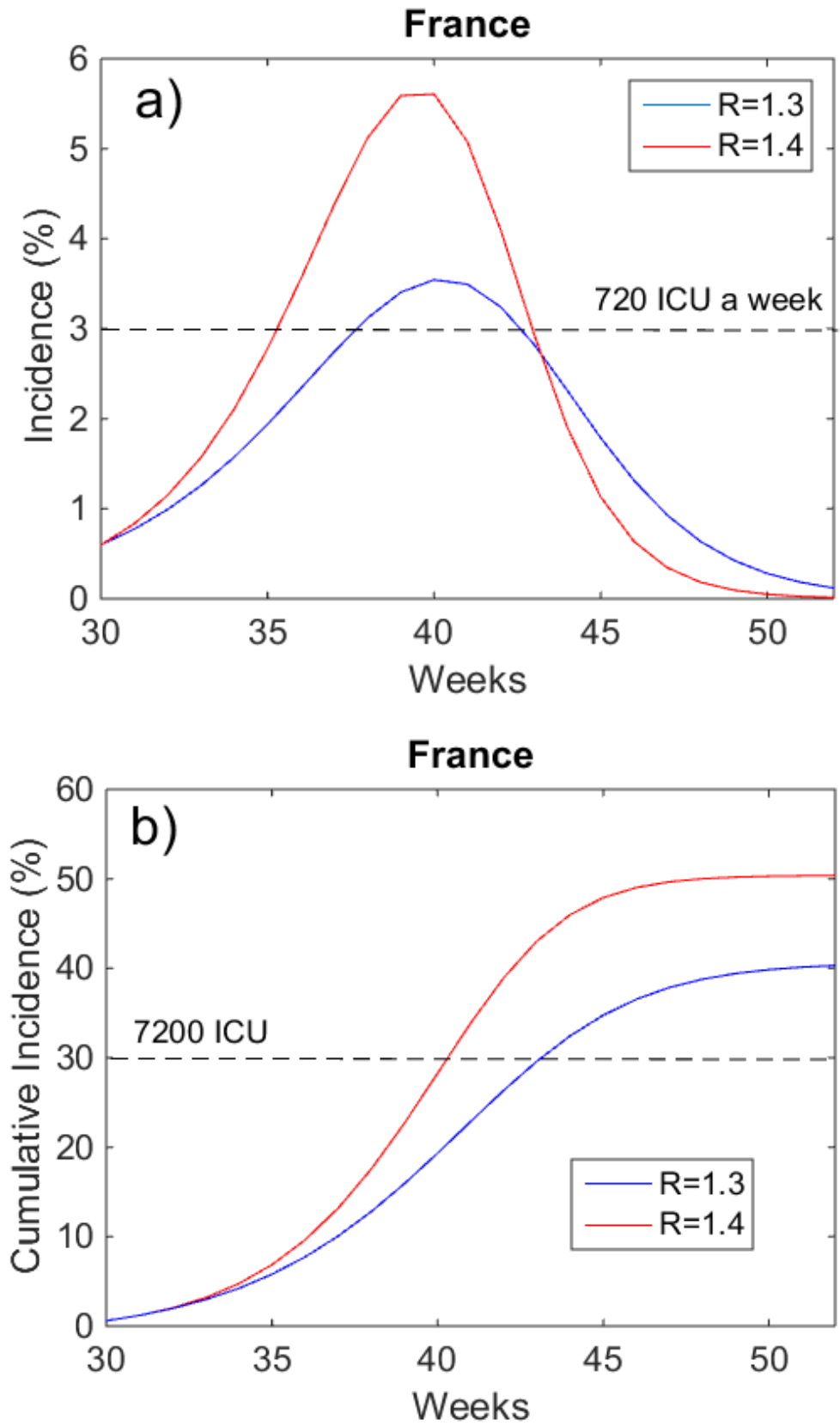

**Figure 1.** France. a) Time evolution of incidence. The horizontal dashed line shows the estimated ICU admissions each weak for an incidence of 3%. b) Time evolution of cumulative incidence for the second wave. The horizontal dashed line shows the estimated ICU admission for a cumulative incidence of 30 %.

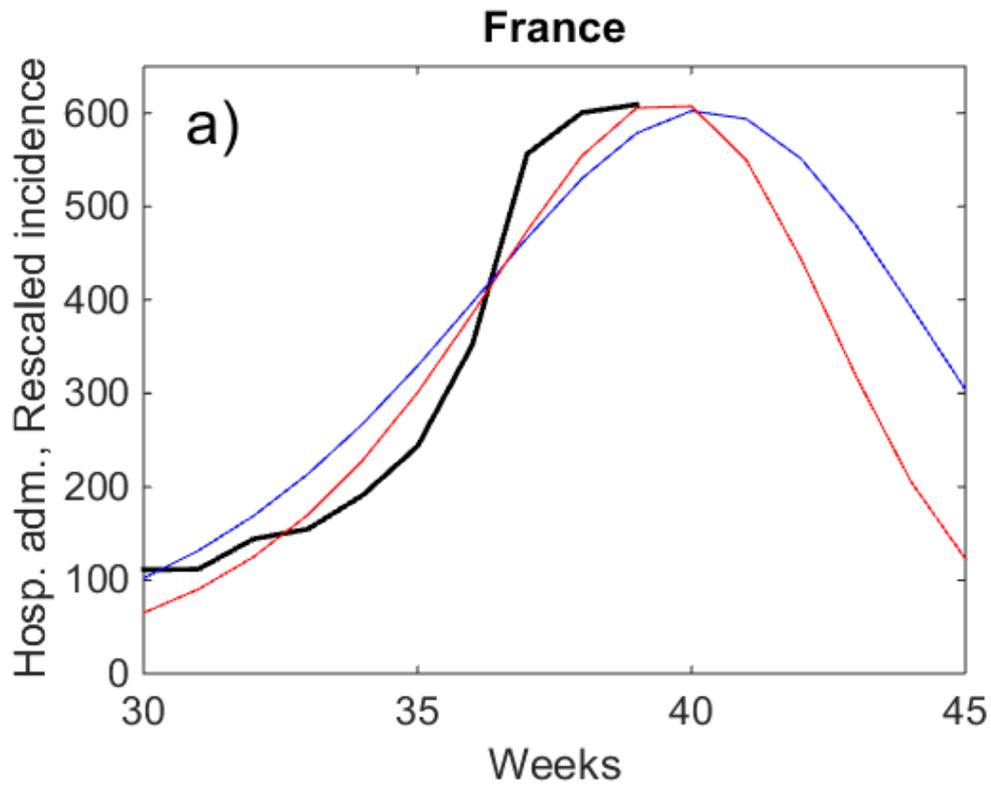

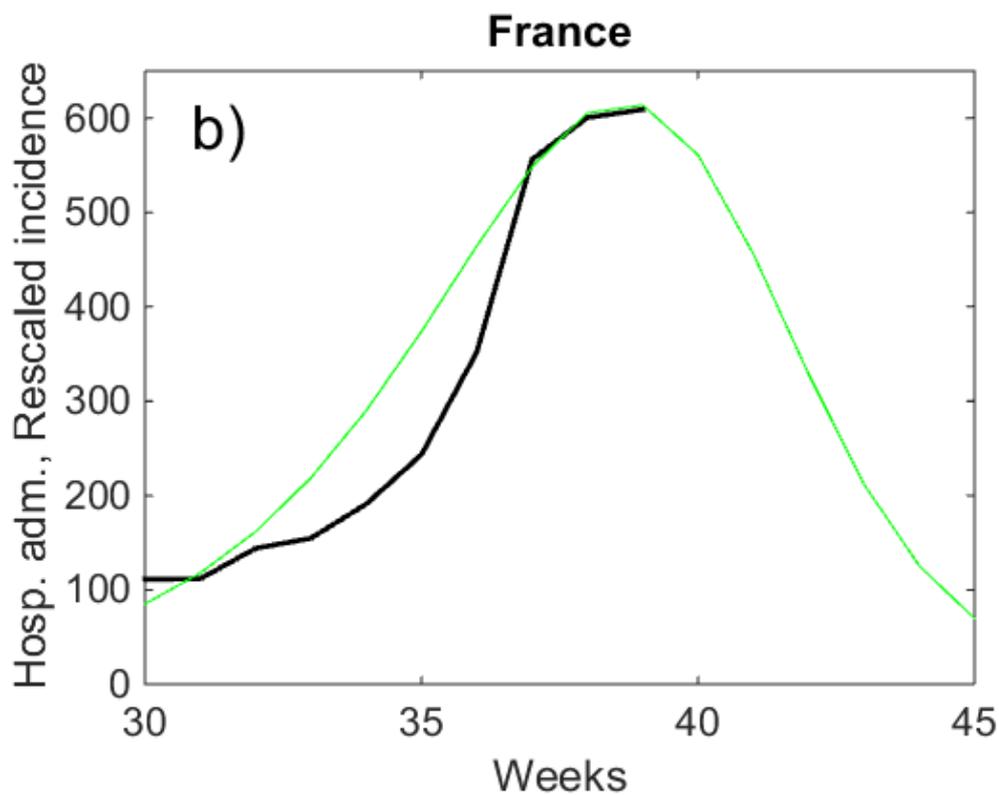

Figure 2: a) Black Hospital admission per day, averaged over one week, and shifted in time by one week. The last point corresponds to the full week 40. Blue: Same as blue curve on fig 1.a but rescaled. Red: Same as blue curve on fig 1.a but rescaled. b) Black Hospital admission per day, averaged over one week, and shifted in time by two weeks. Green: Rescaled incidence computed with $R$=1.4 (Same as red curve), but with $f_s$=60%.

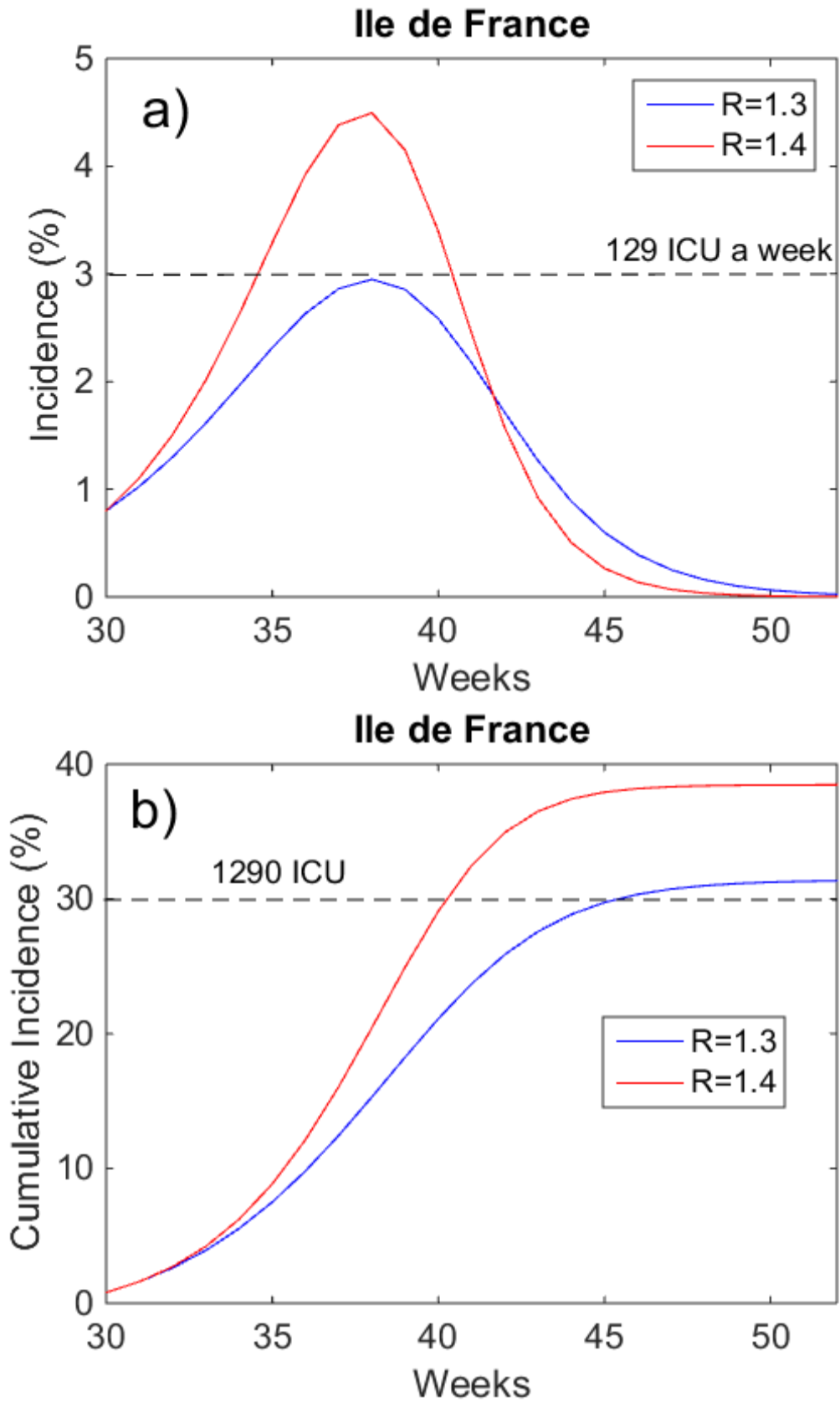

**Figure 3.** Same as Figure 1 for Ile de France (parameters given in table 3).

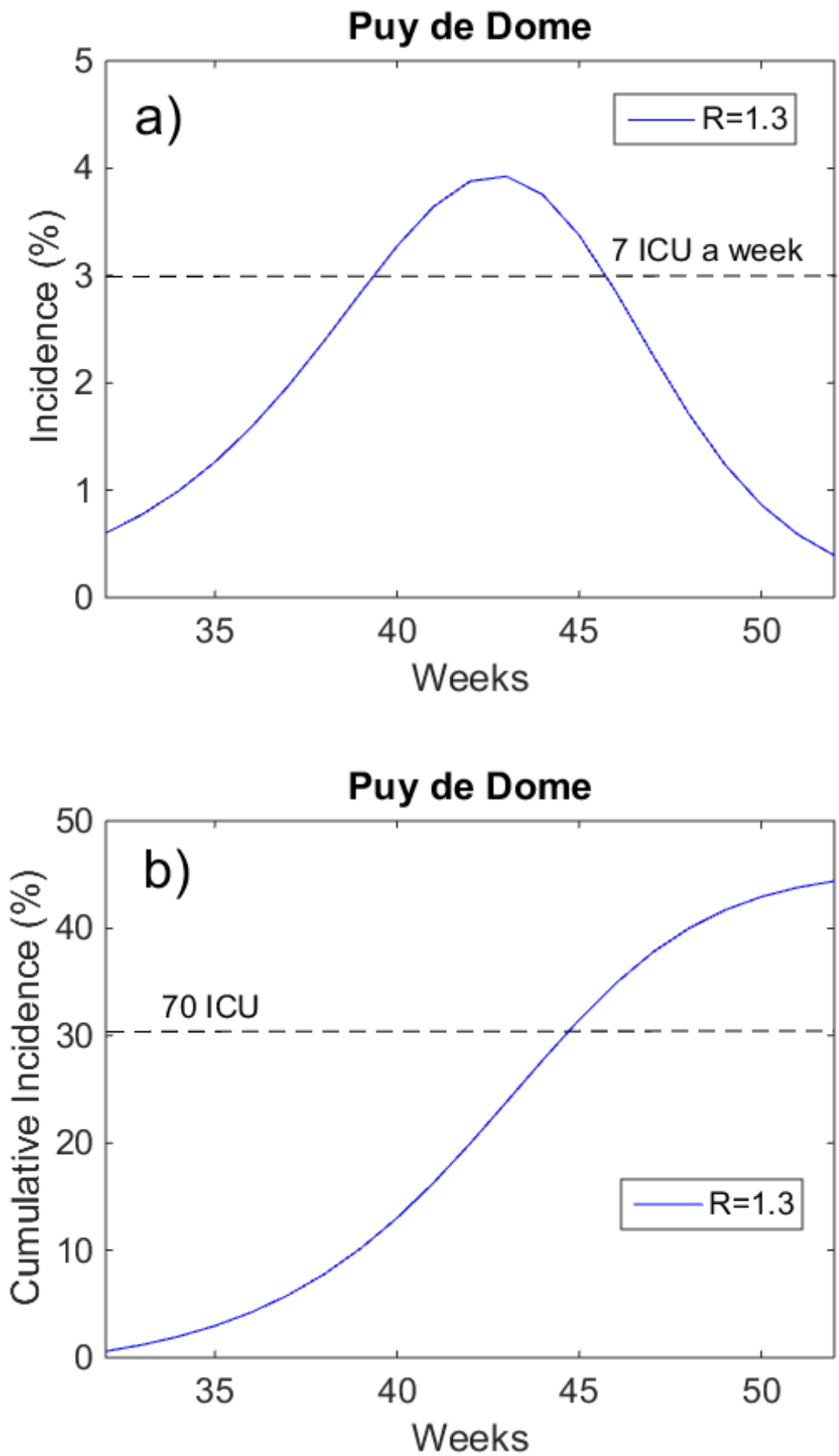

**Figure 4.** Same as Figure 1 for Puy de Dome (parameters given in table 3).

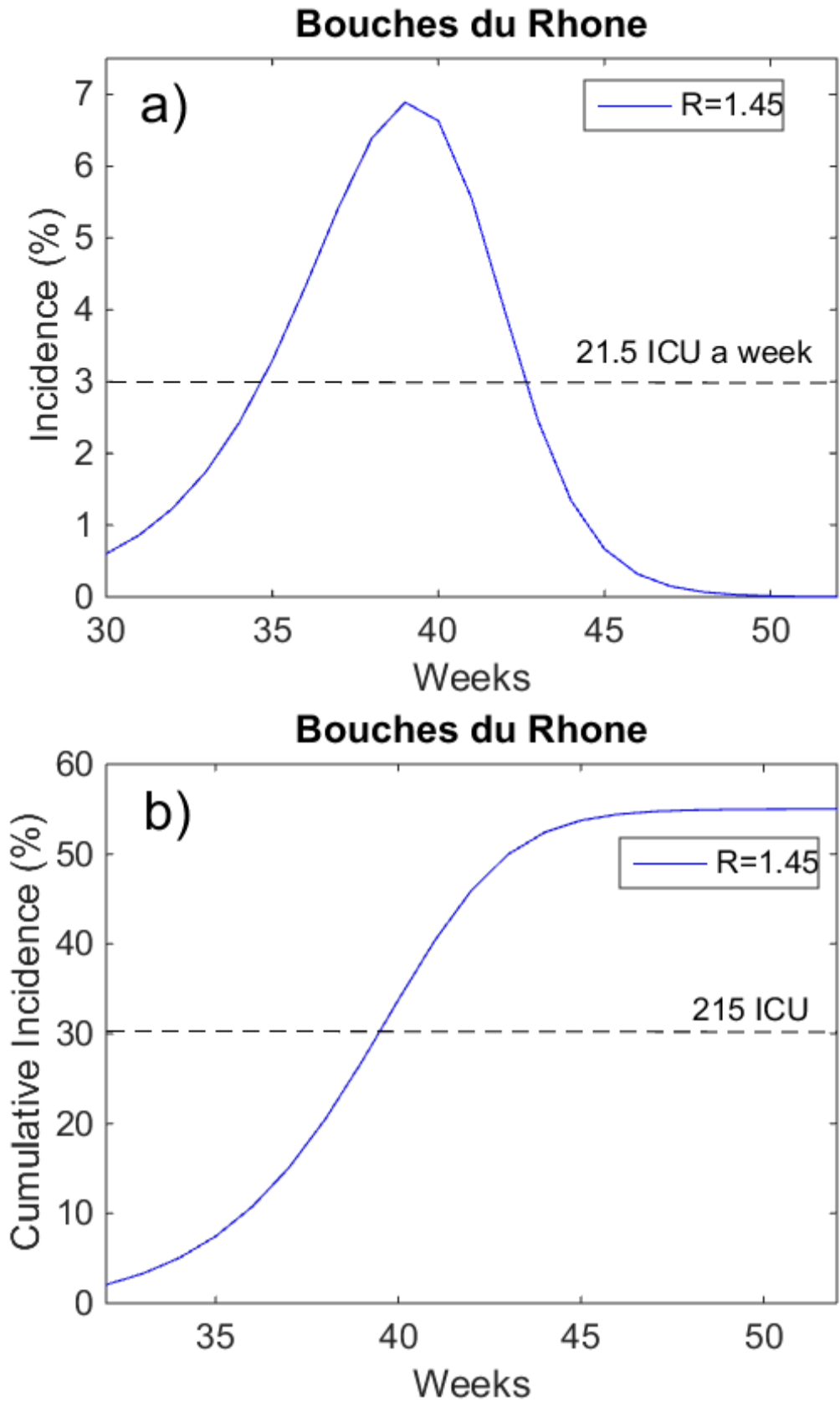

**Figure 5.** Same as Figure 1 for Bouches du Rhone (parameters given in table 3).

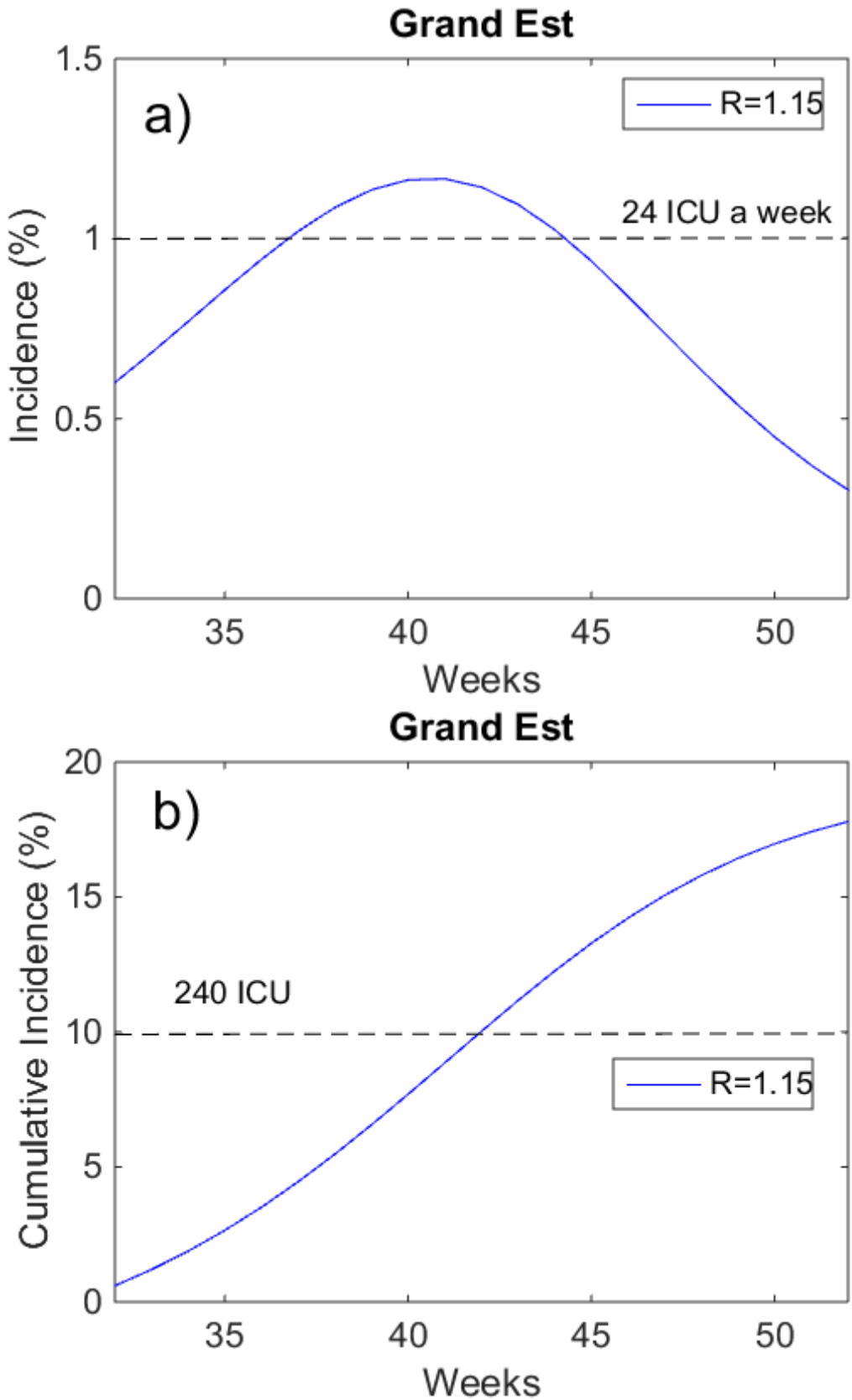

**Figure 6.** Same as Figure 1 for Grand Est (parameters given in table 3).

**VII Discussion-Conclusion**

This work is essentially based on the analysis of Santé Public reports. It enunciates an evidence. Testing a large population of asymptomatic people randomly in a large population allows to determine the real incidence of this population. As a consequence, the number of infected people in France, at a given time, at the beginning of September is not 50.000 but rather 1.6 millions. The level of spreading raises doubts in the efficiency of the large-scale testing strategy which anyway detects only a few percents of infected people. It also raises doubts in the efficiency of the closing of school classes, where a pupil is tested Covid-positive by chance, a procedure judiciously abandoned at the beginning of September. This counting also leads to another evidence, the IFR, the probability of severe forms decayed since March-April by more than a factor 10 and is now about 0.02%-0.027%. Simulations assuming that re-infections are not likely show that the second wave peak should occur at the end of September - beginning of October. In this second wave the occurrence of an infection peak is due to the development of a large enough immunity in the population. The simulated total number of infections during this wave is probably larger than the one (not fully known) related to the first wave, which was controlled by a severe lockdown. However, the expected number of death is 4 times smaller. The very basic simulation tool we use, with essentially an homogeneous constant in time $R$-value do not allow to reproduce precisely all the details of the outbreak evolution but rather gives a general picture which our hypothesis are verified is very likely to occur.

We might also speculate on a possible scenario for the future evolution of the outbreak. It might be indeed that individual immunity is really short term, so that multiple re-infections are possible, but that the severity really decreases, in average, when these re-infections occur. May be the destiny of SARS-COV2 is to end as other coronaviruses, so to be associated to benign infection, but extremely spread and endemic of the earth population. This is purely speculative, but would explain the behavior of a part of experimental data. We also want to insist again that the crucial result and hypothesis we do in the present work, namely the real incidence, and the probability to be infected several times could be verified in quite simple way by analyzing carefully existing data.

We finally want to insist that the general validity of these simulations also rely on keeping the bare transmission rate as it is now (at least of the same order), namely limited by social distancing and restrictions. Indeed if these measures are relaxed now, about 100 % of the population will be infected, and the cumulative incidence of the second wave will be 80 % instead of 40%. ICU admissions will double and become more numerous than during the first wave. Direct covid mortality could reach 18.000 persons. Indirect mortality related to the overwhelmed health system could be very significant. One should note, that while writing these lines, the epidemic is suddenly rising in France, with an observed reproduction rate (so including potential immunity) which suddenly rose from 1 to 1.3 and the situation became critical. This might be associated with a sudden change of weather conditions which became quite suddenly winter like obliging people to go inside, with closed windows. If our estimations are correct and if individual immunity exists, this will modify the conditions for some large enough immunity to be acquired by a few weeks. If not, only increased restrictions could limit the outbreak extension.

**Acknowledgments:** GM thanks Dmitry Solnyshkov for discussions, corrections of the manuscript and technical support. We acknowledge the support of the ANR program "Investissements d'Avenir" through the IDEX-ISITE initiative 16-IDEX-0001 (CAP 20-25).

**Author contribution statement:** GM has coordinated the work. GM implemented all technical aspects. GM, AT, FB and LG wrote the manuscript.

Competing interest statement: The authors declare that they have no conflict of interest.